\documentstyle[prb,twocolumn,aps,rotate,epsf,floats,amssymb,amsmath]{revtex}

\newcommand{\bq}{{\boldsymbol q}}
\newcommand{\bk}{{\boldsymbol k}}
\newcommand{\bl}{{\boldsymbol l}}

\newcommand{\bR}{{\boldsymbol R}}
\newcommand{\bS}{{\boldsymbol S}}
\newcommand{\br}{{\boldsymbol r}}

\newcommand{\ox}{{\overline{x}}}
\newcommand{\oy}{{\overline{y}}}
\newcommand{\oq}{{\overline{q}}}
\newcommand{\om}{{\overline{m}}}
\newcommand{\vF}{{v_{\rm F}}}
\newcommand{\ovF}{{\overline{v}_{\rm F}}}
\newcommand{\ovr}{{\overline{r}}}

\newcommand{\obr}{{\overline{\boldsymbol r}}}

\def\rbx#1{\raisebox{-0.3ex}{$\scriptstyle #1$}}

\begin{document}
\setlength{\unitlength}{1cm}
\renewcommand{\arraystretch}{1.4}

\title{Low temperature electronic properties of Sr$_2$RuO$_4$ I:\\
Microscopic model and normal state properties}  

\author{Ralph Werner\thanks{Present address: Institut f\"ur Theorie der
                  Kondensierten Materie, Universit\"at Karlsruhe, 76128
                  Karlsruhe, Germany.} and V. J. Emery} 
\address{Physics Department, Brookhaven National
Laboratory, Upton, NY 11973-5000, USA} 

\date{\today}

\maketitle

\centerline{Preprint. Typeset using REV\TeX.}

\begin{abstract} 
Starting from the quasi one-dimensional kinetic energy of the $d_{\rm
yz}$ and $d_{\rm zx}$ bands we derive a bosonized description of the
correlated electron system in Sr$_2$RuO$_4$. At intermediate coupling
the magnetic correlations have a quasi one-dimensional component along
the diagonals of the basal plane of the tetragonal unit cell that
accounts for the observed neutron scattering results. Together with
two-dimensional correlations the model consistently accounts for the
normal phase specific heat, cyclotron mass enhancement, static
susceptibility, and Wilson ratio and implies an anomalous high
temperature resistivity.
\end{abstract}


\section{Introduction}

Sr$_2$RuO$_4$  is the first layered transition metal oxide that
exhibits superconductivity in the absence of copper
ions.\cite{MHY+94} The lattice symmetry is tetragonal and
isostructural to La$_2$CuO$_4$ with lattice parameters $a=b=3.87$
{\AA} in the RuO$_2$ plane and $c=12.74$ {\AA} out-of-plane. No
structural instabilities are observed.\cite{BRN+98} The first de
Haas--van Alphen (dHvA) results\cite{MJD+96} and band structure
calculations in local density approximation (LDA)\cite{Oguc95} show
three bands cutting the Fermi level with quasi two-dimensional Fermi
surfaces. They can be mainly associated with the three $t_{2g}$
orbitals of the Ru$^{4+}$ ions,\cite{SCB+96,MRS01} and are consistent
with the metallic properties and the strongly anisotropic transport
along the $c$ axis.\cite{MHY+94}  

The enhanced specific heat and magnetic susceptibility indicate the
presence of significant correlations.\cite{MHY+94} Consistently, 
results from angle resolved photo emission spectroscopy
(ARPES)\cite{PSKT98} and dHvA measurements\cite{MJD+96} suggest a
strong electronic mass renormalization. The material is Fermi liquid
like in a temperature range of $T_{\rm c} < T < 30$
K.\cite{MYH+97,MIM+98,IMK+00} 

The significant correlations in Sr$_2$RuO$_4$, the $S = 1$ moments on
Ru$^{4+}$ impurities in Sr$_2$IrO$_4$,\cite{CBK+94} and ferromagnetic
correlations in SrRuO$_3$ led Rice and Sigrist\cite{RS95} to propose
that the superconducting order parameter has $p$-wave symmetry
promoted by ferromagnetic correlations analogous to $^3$He. The
absence of a change in the Knight shift in the superconducting
phase\cite{IKA+97,IMK+98} supported that notion as well as the
temperature independent magnetic susceptibility\cite{DHM+00} and the
enhanced relaxation time in Muon spin resonance ($\mu$SR)\cite{LFK+98}
at $T\le T_{\rm c} \sim 1.5$ K. A similar proposal was made by
Baskaran based on a comparison with high $T_{\rm c}$ materials and
emphasizing the role of Hund's rule coupling.\cite{Bask96}  

Since then tremendous experimental effort has been made trying to
verify the predicted $p$-wave symmetry of the superconducting order 
parameter. Neither tunneling\cite{LGL+00,MNJ+01} nor thermal
conductivity experiments\cite{TNM+01,TSN+01,ITY+01} or
ac-susceptibility measurements\cite{YAM+01} under different magnetic
field geometries gave conclusive proof of the analogy to $^3$He. No
indication for ferromagnetic correlations has been found in
ARPES,\cite{DLS+00,SDL+01} LDA,\cite{MS99,MPS00} or neutron
scattering\cite{SBB+99,BSB+02} investigations. Furthermore, the
specific heat\cite{NMF+98,NMM99,NMM00} and nuclear quadrupole
resonance (NQR)\cite{IMK+00} are consistent with two-dimensional
gapless fluctuations in the  superconducting phase of Sr$_2$RuO$_4$
which are absent in superfluid $^3$He.

The controversy about the proper description of the electronic
correlations in Sr$_2$RuO$_4$ is reflected most impressively by the
variation of values of the on-site Coulomb repulsion $U$ used in the
mostly perturbative approaches to match experimental results. Examples
are $U \approx 0.42$ eV,\cite{MS97} $0.2$ eV,\cite{LAG00} $1.2$--$1.5$
eV,\cite{LL00} $0.345$ eV,\cite{EMJB02,EMB02} $0.175$ eV,\cite{MTG01}
$2$ eV,\cite{KOAA01} $0.048$ eV.\cite{ALGW01} Comparing these
values of the interaction with the bare Fermi velocity of $\ovF\approx
0.7$ eV$a$ from band structure calculations\cite{MPS00} and
ARPES\cite{SDL+01} points toward an intermediate coupling 
regime. 

In contrast to the effects of the interactions the bare electronic
band structure has been determined unambiguously from
dHvA,\cite{BJM+00,BMJ+02} ARPES,\cite{DLS+00} and x-ray-absorption
measurements\cite{SCB+96} in consistency with LDA
calculations.\cite{MS97,MPS00} The overlap of the electronic wave
functions of the $d_{zx}$ and $d_{yz}$ orbitals is dominantly one
dimensional.\cite{MS97,MPS00,MRS01} The interaction and additional
hopping channels lead the $d_{zx}$ and $d_{yz}$ electrons to hybridized
into two bands. Their Fermi surfaces are referred to as the $\alpha$
and $\beta$ sheets. The electrons in the $d_{xy}$ orbital form the
two-dimensional $\gamma$ sheet.\cite{MS97,MPS00}

Correlations in effective one-dimensional systems show power
law behavior.\cite{SCP98} They are always more singular than
two-dimensional correlations which diverge at most 
logarithmically.\cite{Schu95} Since the kinetic energy of the $d_{zx}$
and $d_{yz}$ electrons is quasi one-dimensional we expect their
correlations to play a dominant role. 

The quasi one-dimensional kinetic energy of the $d_{zx}$ and $d_{yz}$
electrons allows for the bosonized, non-perturbative description of
the low energy electronic excitations. This description is introduced
in Sec.\ \ref{sectionzbands} and its fundamental properties are 
discussed. Section \ref{sectionhybrid} is devoted to the expected
corrections from hybridization effects and the $\gamma$ sheet that
have been neglected in the initial model. The comparison with
experimental results in Sec.\ \ref{sectionapplic} reveals the
qualitative and quantitative consistency of the model within the
framework of its applicability. A comparative discussion of
alternative perturbative approaches is included [Sec.\
\ref{sectionRPA}].    

The present paper is part I of a series of three. Part II (Ref.\
\onlinecite{Wern02b}) is devoted to the superconducting phase. The
in-plane correlations are described via the model derived herein. The
inter-plane pair-correlations are enhanced as a consequence of the body
centered crystal structure and can be treated mean-field like. 

Part III (Ref.\ \onlinecite{Wern02c}) consistently explains the
experimentally observed unconventional transitions under magnetic
fields based on the model derived here and in part II.


\section{Subsystem of \lowercase{$d_{zx}$} and \lowercase{$d_{yz}$}
bands}\label{sectionzbands}  

The band structure as determined from dHvA,\cite{BJM+00}
ARPES,\cite{DLS+00} and x-ray-absorption measurements\cite{SCB+96} as
well as LDA calculations\cite{MS97,MPS00} together with the
anticipated intermediate interactions suggest a three band Hubbard
Hamiltonian as the generic model. 
\begin{equation}\label{genericH}
H=\sum_{{\bl},{\bl'}\atop\nu,\sigma} t^{\nu,\nu'}_{{\bl},{\bl}'} 
         c^{\dagger}_{{\bl},\nu,\sigma} 
         c^{\phantom{\dagger}}_{{\bl}',\nu,\sigma}
+ \sum_{{\bl},\nu,\sigma\atop\nu',\sigma'} U^{\nu,\nu'}_{\sigma,\sigma'}\,
        n_{{\bl},\nu,\sigma} n_{{\bl},\nu',\sigma'}.
\end{equation}
In this notation the electron creation and annihilation operators 
are $c^{\dagger}_{{\bl},\nu,\sigma}$ and
$c^{\phantom{\dagger}}_{{\bl},\nu,\sigma}$ for orbital $\nu$ with
spin $\sigma$ on site $\bl$, $n_{{\bl},\nu,\sigma}$ is the usual
electronic density operator, $t^{\nu,\nu'}_{{\bl},{\bl}'}$ is the
hopping matrix element between site ${\bl}$ and ${\bl}'$, and
$U^{\nu,\nu'}_{\sigma,\sigma'}$ is the on-site Coulomb repulsion. 

As discussed in the introduction we expect the interesting low
temperature physics to be dominated by the quasi one-dimensional
$d_{zx}$ ($\nu=x$) and $d_{yz}$ ($\nu=y$) bands. We retain here only
the dominant hopping amplitudes $t_0 = t^{x,x}_{{\bl},{\bl} + \hat{x}}
= t^{y,y}_{{\bl},{\bl} + \hat{y}}$ and discuss effects from the
hybridization of the bands later in Sec.\ \ref{sectionhybrid}. The
continuum representation is introduced via $c_{{\bl},\nu,\sigma} \to
\psi_{\nu,\sigma}(\br)$ with $\rho_{\nu,\sigma}(\br) =
\psi^\dagger_{\nu,\sigma}(\br)
\psi^{\phantom\dagger}_{\nu,\sigma}(\br)$. The bands are linearized
with Fermi velocity $\vF \approx \sqrt{3}\ t_0$. Note that the
``velocities'' in the present paper define the kinetic energy scales,
i.e., $\vF = \ovF/a$ ($\hbar\equiv 1$).   
\begin{eqnarray}\label{1DH}
H_{\rm 2D} &=& \lim_{a\to 0 \atop L\to \infty} 
\sum_{\nu,\sigma}\int_{-L}^L d^2r\ \bigg[
   i\ v_{\rm F}\ \psi^{\dagger}_{\nu,\sigma}(\br)\,\partial_\nu\, 
         \psi^{\phantom{\dagger}}_{\nu,\sigma}(\br)\nonumber\\&&\hspace{0ex}
+ \frac{1}{4}\sum_{\nu',\sigma'} 
   \rho_{\nu,\sigma}(\br) \Big(U_0\ \sigma^0_{\nu,\nu'}
             \sigma^x_{\sigma,\sigma'}
        +U_1\ \sigma^x_{\nu,\nu'}
             \sigma^x_{\sigma,\sigma'} 
\nonumber\\&&\hspace{18ex}
        +\ U_2\ \sigma^x_{\nu,\nu'}
             \sigma^0_{\sigma,\sigma'}\Big) \rho_{\nu',\sigma'}(\br)
\bigg]
\end{eqnarray}
Here $\sigma^{x,y,z,0}_{a,a'}$ denote the Pauli matrices with
$\sigma^{0}_{a,a'} = (\sigma^{z}_{a,a'})^2$ and $2L=L_x=L_y$ is the
linear dimension of the system. We limit the description here to the
RuO$_2$ planes and generalize when necessary to a three-dimensional
array of planes. The intra-orbital Coulomb repulsion is larger than
the inter-orbital repulsion, i.e., $U^{\nu = \nu'}_{\sigma\neq\sigma'}
= U_0 > U_1 = U^{\nu \neq \nu'}_{\sigma \neq \sigma'}$ and $U_0 > U_2
= U^{\nu \neq \nu'}_{\sigma = \sigma'}$. 

Hund's rule coupling lowers the inter-orbital Coulomb repulsion for
electrons in a spin-triplet configuration with respect to the spin
singlets. The full treatment of the involved exchange interaction
terms within the framework of the bosonization approach discussed in
Sec.\ \ref{sectionbose} is rather involved and only possible in
approximations. A qualitative study of the effect of Hund's rule
coupling is possible by setting $U_1>U_2$ and neglecting the exchange
terms. The expected corrections due to exchange and other terms are
discussed closer in Sec.\ \ref{sectionHund}.

The Hamiltonian Eq.\ (\ref{1DH}) is SU(2) invariant both in the spin 
sector and in the orbital sector yielding an effective
SU(2)$\otimes$SU(2) symmetry. The orbital degrees of freedom are 
sometimes referred to as electron flavors.\cite{EK92,GNT98}

\subsection{Bosonization}\label{sectionbose}

In Eq.\ (\ref{1DH}) the $d_{zx}$ fields only have a kinetic energy 
contribution along the $x$ direction while the $d_{yz}$ fields only
have a kinetic energy along $y$. We can thus switch to the chiral
representation\cite{Frad91} 
\begin{equation}\label{chiral}
\psi_{\nu,\sigma}(\br)=
        R_{\nu,\sigma}(\br)\,{\rm e}^{ik_{\rm F}r_\nu} +
        L_{\nu,\sigma}(\br)\,{\rm e}^{-ik_{\rm F}r_\nu} 
\end{equation}
for each orbital degree of freedom $\nu$ at each position
$r_{\nu'\neq\nu}$ transverse to the propagation. The right
($R_{\nu,\sigma}$) and left ($L_{\nu,\sigma}$) moving fermions of each
species can now be bosonized.\cite{EK92,SCP98} 
\begin{eqnarray}\label{bosonizationR}
 R_{\nu,\sigma}(\br)&=&\lim_{a\to 0}
\frac{\eta^R_{\nu,\sigma}(r_{\nu'})}{\sqrt{2\pi a}}\
          {\rm e}^{-i\sqrt{\pi}[\theta_{\nu,\sigma}(\br)+
                                \phi_{\nu,\sigma}(\br)]}
\\              \label{bosonizationL}
 L_{\nu,\sigma}(\br)&=&\lim_{a\to 0}
\frac{\eta^L_{\nu,\sigma}(r_{\nu'})}{\sqrt{2\pi a}}\
          {\rm e}^{-i\sqrt{\pi}[\theta_{\nu,\sigma}(\br)-
                                \phi_{\nu,\sigma}(\br)]}
\end{eqnarray}
Here $\phi_{\nu,\sigma}(\br)$ are the Bose fields with their conjugate
momenta $\Pi_{\nu,\sigma}(\br)=\partial_\nu \theta_{\nu,\sigma}(\br)$
which satisfy the commutation relation
\begin{equation}\label{bosecommute}
\left[\phi_{\nu,\sigma}(\br),\Pi_{\nu',\sigma'}(\br')\right] =
i\,\delta_{\nu,\nu'}\,\delta_{\sigma,\sigma'}\,\delta(\br-\br').
\end{equation}
The Klein factors\cite{Sene99} $\eta_{\nu,\sigma}(r_{\nu'})$ assure
the proper commutation relation between the different fermion species,
$a$ is a short range cut off associated with the in-plane lattice
constant. The bosonized Hamiltonian can be written  
as 
\begin{eqnarray}\label{bosonH}
H_{\rm 2D} &=& \lim_{a\to 0 \atop L\to \infty} \sum_{\sigma\neq\sigma'}
          \sum_{\nu\neq\nu'} \int_{-L}^L\!\! d^2r\!
\ \bigg[\frac{\vF}{2}\left(\Pi^2_{\nu,\sigma}  + \left(\partial_\nu\, 
         \phi_{\nu,\sigma} \right)^2\right)
\nonumber\\&&\hspace{-4ex}
+\, \frac{U_0}{4} \left( [\partial_\nu\phi_{\nu,\sigma} ]
                       [\partial_{\nu}\phi_{\nu,\sigma'} ] 
         + \frac{\cos \sqrt{4\pi}(\phi_{\nu,\sigma}  - 
                            \phi_{\nu,\sigma'})}{(2\pi a)^2} \right)
\nonumber\\&&\hspace{-4ex}
+\, \frac{U_1}{4} [\partial_\nu\phi_{\nu,\sigma} ]
                       [\partial_{\nu'}\phi_{\nu',\sigma'} ] 
  + \frac{U_1}{4 (2\pi a)^{2}} 
\nonumber\\[-1ex]&&\hspace{0ex}\times 
         \Big(\cos \big[\sqrt{4\pi}(\phi_{\nu,\sigma}  - 
                            \phi_{\nu',\sigma'}) 
                         - 2k_{\rm F}(r_\nu - r_{\nu'})\big]
\nonumber\\[0ex]&&\hspace{2ex}+
               \cos \big[\sqrt{4\pi}(\phi_{\nu,\sigma}  + 
                            \phi_{\nu',\sigma'}) 
                         - 2k_{\rm F}(r_\nu + r_{\nu'})\big] \Big)
 \nonumber\\&&\hspace{-4ex}
+\, \frac{U_2}{4} [\partial_\nu\phi_{\nu,\sigma} ]
                       [\partial_{\nu'}\phi_{\nu',\sigma} ] 
 + \frac{U_2}{4 (2\pi a)^{2}} 
\nonumber\\[-1ex]&&\hspace{0ex}\times 
         \Big(\cos \big[\sqrt{4\pi}(\phi_{\nu,\sigma}  - 
                            \phi_{\nu',\sigma}) 
                         - 2k_{\rm F}(r_\nu - r_{\nu'})\big] 
\nonumber\\[0ex]&&\hspace{2ex}+ 
               \cos \big[\sqrt{4\pi}(\phi_{\nu,\sigma}  + 
                            \phi_{\nu',\sigma}) 
                         - 2k_{\rm F}(r_\nu + r_{\nu'})\big]\Big)
\bigg]\!.\!\!
\nonumber\\&&
\end{eqnarray}

The standard approach to separate spin and charge degrees of freedom
in Eq.\ (\ref{bosonH}) is to introduce charge and magnetic fields for
each flavor via 
\begin{eqnarray} 
\varphi_{\nu,{\rm m}}(\br)&=&[\phi_{\nu,\uparrow}(\br) - 
         \phi_{\nu,\downarrow}(\br)] / \sqrt{2}\,,
\\
\varphi_{\nu,{\rm c}}(\br)&=&[\phi_{\nu,\uparrow}(\br) + 
         \phi_{\nu,\downarrow}(\br)]/\sqrt{2}\,,
\end{eqnarray}
respectively. $\Pi_{\nu,\rm m}$ and $\Pi_{\nu,\rm c}$ are the corresponding
conjugate momenta. The bilinear part of the Hamiltonian $H_{\rm 2D}$
is composed of the charge and (magnetic) part  
\begin{eqnarray}\label{Hc(m)}
H_{\rm c (m)}&=&
 \frac{1}{2}\sum_\nu \int_{}^{} d^2r
    \Big[\vF\, \Pi^2_{\nu,{\rm c (m)}} 
\nonumber\\&&\hspace{2ex} +\ \sum_{\nu'}
         (\partial_\nu \varphi_{\nu,{\rm c (m)}}) 
                    V^{(\rm c (m))}_{\nu,\nu'}
         (\partial_{\nu'} \varphi_{\nu',{\rm c (m)}})
\Big]\,.
\end{eqnarray}
The matrix elements for the charge (magnetic) part are given by 
$
V^{(\rm c(m))}_{x,x} = V^{(\rm c(m))}_{y,y} = 
            \vF + (-)U_0 $
and
$
V^{(\rm c(m))}_{x,y} = V^{(\rm c(m))}_{y,x} = 
            U_1 + (-)U_2
$.
Equation (\ref{Hc(m)}) is the Hamiltonian of a crossed {\em sliding 
Luttinger liquid} studied in Refs.\ \onlinecite{MKL01} and
\onlinecite{KKGA02}. The authors find no significant change in the
decay of the low energy correlations with respect to the
one-dimensional case where $V^{(\rm c(m))}_{x,y} = 0$. A perturbative
treatment suggests that the inclusion of the interaction term in Eq.\
(\ref{bosonH}) leads to two-dimensional correlations which still decay
algebraically.\cite{MKL01} In the absence of Hund's rule coupling,
where $U_2=U_1$, the magnetic sector fields $\varphi_{x,{\rm m}}$ and
$\varphi_{y,{\rm m}}$ are decoupled. Similar models are obtained for
coupled Luttinger liquids.\cite{DL72,Emer79,EFK+00}   

To study the qualitative properties of the model defined by Eq.\
(\ref{bosonH}) with parameters relevant for Sr$_2$RuO$_4$ it proves
useful to use the symmetry of the orbital degrees of  
freedom. We introduce charge ($\mu=\rho$), spin ($\mu={\rm s}$),
flavor ($\mu={\rm f}$), and spin-flavor ($\mu={\rm sf}$) fields via
the canonical transformation
\begin{eqnarray}\label{phimu}
\phi_\mu(\br)&=&\frac{1}{2}\sum\nolimits_{\nu,\sigma} 
       \sigma^a_{\nu,\nu}\sigma^b_{\sigma,\sigma}\
              \phi_{\nu,\sigma}(\br)\,,
\\              \label{Pimu}
\Pi_\mu(\br)&=&\frac{1}{2}\sum\nolimits_{\nu,\sigma} 
       \sigma^a_{\nu,\nu}\sigma^b_{\sigma,\sigma}\
              \Pi_{\nu,\sigma}(\br)\,.
\end{eqnarray}
The matrices are $(a,b)=\{(0,0);(0,z);(z,0);(z,z)\}$ for
$\mu=\{\rho;{\rm s};{\rm f};{\rm sf}\}$, respectively. The fields
$\phi_\mu(\br)$ are identical to those of the resonant-level
model used to describe the two-channel Kondo problem.\cite{EK92}
The fields are simple linear combinations of the charge and magnetic
fields, e.g.,  
\begin{eqnarray}\label{phis}
\phi_{\rm s}(\br) & = & \left( \varphi_{x,{\rm
m}}+\varphi_{y,{\rm m}} \right)/\sqrt{2} 
\\              \label{phisf}
\phi_{\rm sf}(\br) & = & \left(  \varphi_{x,{\rm 
m}}-\varphi_{y,{\rm m}} \right)/ \sqrt{2}\,.
\end{eqnarray} 
Note also that the charge and spin sector fields
are symmetry related via the reflection $y\to -y \Rightarrow R_{y,\sigma} 
\leftrightarrow L_{y,\sigma} \Rightarrow \phi_{\rm f} \leftrightarrow
\phi_{\rho}$ and $\phi_{\rm s} \leftrightarrow \phi_{\rm sf}$.

The representation can be simplified by introducing the variables
$\ox=\frac{1}{\sqrt{2}}(x+y)$ and $\oy=\frac{1}{\sqrt{2}}(x-y)$ with
$\obr=(\ox,\oy)^\dagger$. The charge Hamiltonian in Eq.\ (\ref{Hc(m)})
becomes  
\begin{eqnarray}\label{Hnullc}
H_{\rm c} &=& \frac{1}{2} \int_{}^{}\!\! d^2\ovr\,
    \Big\{\vF \left(\Pi^2_{\rho} + \Pi^2_{\rm f}\right) 
\nonumber\\&&\hspace{0ex}
 +\ V_{\rm c} \left[ \partial_{\ox} \phi_{\rho}
    + \partial_{\oy} \phi_{\rm f} \right]^2
 +\ \overline{V}_{\rm c} \left[ \partial_{\oy} \phi_{\rho}
    + \partial_{\ox} \phi_{\rm f} \right]^2
\Big\}\, ,
\end{eqnarray} 
while the magnetic Hamiltonian is 
\begin{eqnarray}\label{Hnullm}
H_{\rm m} &=& \frac{1}{2} \int_{}^{}\!\! d^2\ovr\,
    \Big\{\vF\left(\Pi^2_{\rm s} + \Pi^2_{\rm sf}\right) 
\nonumber\\&&\hspace{-2ex}
 +\ V_{\rm m} \left[ \partial_{\ox} \phi_{\rm s}
    + \partial_{\oy} \phi_{\rm sf} \right]^2
 +\ \overline{V}_{\rm m} \left[ \partial_{\oy} \phi_{\rm s}
    + \partial_{\ox} \phi_{\rm sf} \right]^2
\Big\}\, .
\end{eqnarray} 
The energies are 
\begin{eqnarray} \label{Vc}
V_{\rm c} &=& \vF+U_0+(U_1+U_2),
\\               \label{oVc}
\overline{V}_{\rm c} &=& \vF+U_0-(U_1+U_2),
\\               \label{Vm}
V_{\rm m} &=& \vF-U_0+(U_1-U_2),
\\               \label{oVm}
\overline{V}_{\rm m} &=& \vF-U_0-(U_1-U_2).
\end{eqnarray}
Applying Eqs.\ (\ref{phimu}) and (\ref{Pimu}) the interaction term in
the Hamiltonian Eq.\ (\ref{bosonH}) factorizes into contributions of
the four spin, charge, flavor, and spin-flavor degrees of freedom.
\begin{eqnarray}\label{Hintz}
H_{\rm int}&=& \frac{U_0}{(2\pi a)^2} \int_{}^{} d^2\ovr
    \cos \sqrt{4\pi}\phi_{\rm s}(\obr)\, 
                     \cos \sqrt{4\pi}\phi_{\rm sf}(\obr)
\nonumber\\&+& \frac{1}{(2\pi a)^2} \int_{}^{} d^2\ovr\
\cos \big[\sqrt{4\pi}\phi_{\rm f}(\obr) - 2\sqrt{2} k_{\rm F} \oy\big]
\nonumber\\[-0ex]&&\hspace{4ex}\times
 \Big(U_1\cos \sqrt{4\pi}\phi_{\rm s}(\obr) + 
                    U_2\cos \sqrt{4\pi}\phi_{\rm sf}(\obr)\Big)
\nonumber\\&+& \frac{1}{(2\pi a)^2} \int_{}^{} d^2\ovr\
\cos \big[\sqrt{4\pi}\phi_{\rho}(\obr) - 2\sqrt{2} k_{\rm F} \ox\big]
\nonumber\\[-0ex]&&\hspace{4ex}\times
 \Big(U_2\cos \sqrt{4\pi}\phi_{\rm s}(\obr) + 
                    U_1\cos \sqrt{4\pi}\phi_{\rm sf}(\obr)\Big)
\nonumber\\[-0ex]&&
\end{eqnarray}
The limit $a\to 0$ and $L\to \infty$ is understood. The total
Hamiltonian of the $d_{zx}$-$d_{yz}$ subsystem is $H_{\rm 2D} = H_{\rm
c} + H_{\rm m} + H_{\rm int}$.  

No spin or charge density wave instabilities are observed in
Sr$_2$RuO$_4$.\cite{SBB+99} The values for the on-site Coulomb
repulsions discussed in the
literature\cite{MS97,LAG00,LL00,EMJB02,KOAA01,MTG01,ALGW01} point
toward an intermediate coupling regime if compared to the bare Fermi
velocity of $\vF\approx 0.7$ eV from band structure
calculations.\cite{MPS00} For repulsive interactions the operators in
Eq.\ (\ref{Hintz}) have been shown to be marginally irrelevant both 
in one and two dimensions.\cite{SCP98} The physical properties are
therefore determined by $H_{\rm c}$ and $H_{\rm m}$ with quantitative
corrections from $H_{\rm int}$. Corrections from hybridization terms
not included in the bosonized model are discussed in
Sec. \ref{sectionhybrid}. 

Note that Eqs.\ (\ref{Hnullc}) through (\ref{Hintz}) are still explicitly
invariant under the reflection $x\leftrightarrow y$ which is equivalent
to $\ox\to \ox$, $\oy \to -\oy$, $\phi_{\rm f} \to -\phi_{\rm f}$, and
$\phi_{\rm sf} \to -\phi_{\rm sf}$. The same applies for $y\to -y$
where $\ox \leftrightarrow \oy$, $R_{y,\sigma} \leftrightarrow
L_{y,\sigma}$, $\phi_{\rm f} \leftrightarrow \phi_{\rho}$ and
$\phi_{\rm s} \leftrightarrow \phi_{\rm sf}$.

\subsection{Effective one-dimensional model}\label{sectiononeeff}

In the intermediate coupling regime the model defined by Eqs.\
(\ref{Hnullc}) through (\ref{Hintz}) exhibits a number of singular
points for $V_{\rm c(m)}=0$ or $\overline{V}_{\rm c(m)}=0$. For $U_0 <
\vF$ and $U_0 > U_1 > U_2$ the relevant limit is $\overline{V}_{\rm m}
\to 0$. Then the magnetic Hamiltonian Eq.\ (\ref{Hnullm}) only has
terms in $\partial_{\ox} \phi_{\rm s}$ and $\partial_{\oy} \phi_{\rm
sf}$. 
\begin{equation}\label{Hnullom}
H_{\rm \om} = \frac{1}{2} \int_{}^{}\!\! d^2\ovr\,
    \Big\{\vF\left(\Pi^2_{\rm s} + \Pi^2_{\rm sf}\right) 
 + V_{\rm m} \left[ \partial_{\ox} \phi_{\rm s}
    + \partial_{\oy} \phi_{\rm sf} \right]^2
\Big\}\,.
\end{equation} 

The representation of the spin and spin-flavor fields introduced in
Sec.\ \ref{sectionbose} has the property that along $x=y$ one finds
$\partial_{\oy} \phi_{\rm s} = \partial_{\ox} \phi_{\rm sf} = 0$. This
becomes obvious from Eqs.\ (\ref{phis}) and (\ref{phisf}) together
with the symmetry\cite{Stablequote} implied relation $\partial_{y}
\varphi_{y,{\rm m}}\big|_{x=y} = \partial_{x} \varphi_{x,{\rm
    m}}\big|_{x=y}$. Thus, the fields $\phi_{\rm s}$ and $\phi_{\rm
  sf}$ indeed depend only on $\oy$ and $\ox$, respectively, as implied
by Eq.\ (\ref{Hnullom}) and $H_{\rm \om}$ is effectively one
dimensional. Since the spin and the spin-flavor channel are symmetry
related the one-dimensional correlations in Eq.\ (\ref{Hnullom}) can
be effectively described by     
\begin{equation}\label{effH}
H_{{\rm eff}}=\frac{v_{\rm eff} L}{2}\int_{}^{} d\ox
    \left[K_{\rm eff}^{}\Pi^2_{\rm eff} + K_{\rm eff}^{-1}          
                         (\partial_\ox \phi_{\rm eff})^2\right]\,. 
\end{equation}
Note that the coupling term $(\partial_{\ox} \phi_{\rm s})
(\partial_{\oy} \phi_{\rm sf})$ in Eq.\ (\ref{Hnullom}) can be
eliminated by a mean-field decoupling and subsequent sliding
transformation.\cite{EFK+00} This approximation is less severe than it
may appear at first sight since along $x=y$ the $\ox$ dependence of
$\phi_{\rm sf}$ and the $\oy$ dependence of $\phi_{\rm s}$ can be
neglected. Consequently, along $x=y$, $(\partial_{\ox} \phi_{\rm s})$
is just a $y$-independent constant with respect to $(\partial_{\oy}
\phi_{\rm sf})$ and vice versa. 

The Luttinger liquid parameter $K_{\rm eff}$ and the velocity $v_{\rm
eff}$ are effective parameters of the theory\cite{Hrenormquote} but
can be associated with 
\begin{equation}\label{effK}
K^{-1}_{\rm eff} \sim \sqrt{V_{\rm m}/\vF}
\end{equation}
and 
\begin{equation}\label{effv}
v_{\rm eff} \sim \sqrt{V_{\rm m}\, \vF}\,.
\end{equation}

The model Eq.\ (\ref{Hnullom}) is explicitly invariant under the
transformation $y\to -y$, where $\ox \leftrightarrow \oy$ and
$\phi_{\rm s} \leftrightarrow \phi_{\rm sf}$. The effective
one-dimensional model Eq.\ (\ref{effH}) describes the quasi
one-dimensional magnetic correlations along both in-plane diagonals of
the tetragonal unit cell. 

In realistic systems $\overline{V}_{\rm m} > 0$ and the effective
model is applicable only for sufficiently large temperatures
$T>\overline{V}_{\rm m}$. A close discussion of the values appropriate 
for Sr$_2$RuO$_4$ is given in Sec.\ \ref{sectionapplic}.  

The total Hamiltonian of the low energy in-plane $d_{zx}$-$d_{yz}$
correlations is given by $H_{\rm 2D}=H_{\rm c}+H_{\rm eff}+H_{\rm
int}$ in Eqs.\ (\ref{Hnullc}), (\ref{effH}), and (\ref{Hintz}). The
interaction term $H_{\rm int}$ can be neglected when appropriately
rescaling the parameters $K_{\rm eff}\to K^*_{\rm eff}$, $V_{\rm c}\to
V_{\rm c}^*$, and $\overline{V}_{\rm c}\to \overline{V}_{\rm c}^*$. 

After eliminating the term  $\sim (\partial_{\ox} \phi_{\rm s})
(\partial_{\oy} \phi_{\rm sf})$ in Eq.\ (\ref{effH}) by a sliding
transformation the magnetic Hamiltonian can be written as a
superposition  $H_{\rm m} \approx H_{\rm s} + H_{\rm sf} \approx
H_{\rm eff}$ where $H_{\rm s}$ and $H_{\rm sf}$ are obtained from Eq.\
(\ref{effH}) by replacing (${\rm eff} \to {\rm s}$) and (${\rm eff}
\to {\rm sf}$, $\ox\to\oy$), respectively. Note that here $v_{\rm s} =
v_{\rm sf} = v_{\rm eff}/2$. In the case of SU(2) symmetry in the spin
subspace the interaction term Eq.\ (\ref{Hintz}) yields a rescaled
$K_{\mu}\to K_{\mu}^*=1$. In Ref.\ \onlinecite{Wern02b} it is
suggested that the spins of the electrons in the $d_{zx}$--$d_{yz}$
subsystem are in an easy-plane configuration which implies $K^*_{\rm
s} < 1$ and $K^*_{\rm sf} \le 1$.

\subsection{Comment on Hund's rule coupling}\label{sectionHund}

The treatment of the full SU(2) invariant Hund's rule coupling
term\cite{Bask96} $J_{\rm H} \bS_\nu \bS_{\nu'\neq\nu}$ is difficult
in the bosonized model. The presence of SU(2) symmetry breaking
Dzyaloshinskii--Moriya interactions is obvious from the observed
anisotropies of the static susceptibilities\cite{IKA+97,MYH+97} and is
consistent with the expected presence of spin-orbit coupling in the Ru
4$d$ orbitals.\cite{HS63,NS00,EMB02} Additional corrections are expected
from Kaplan--Shekhtman--Entin-Wohlman--Aharony terms.\cite{ZMT+98} 

The crucial physical implication of Hund's rule coupling is that it
couples the magnetic degrees of freedom of the different orbitals. The
model introduced in Sec.\ \ref{sectionbose} incorporates this effect
qualitatively as becomes apparent from Eq.\ (\ref{Hc(m)}). Notably is
$V_{\rm m} \neq \overline{V}_{\rm m}$ only for $U_1 \neq U_2$ and
consequently can the quasi one-dimensional model discussed in Sec.\
\ref{sectiononeeff} only be found in the presence of Hund's rule
coupling. The quantitative justification of the model is shown
phenomenologically in Sec.\ \ref{sectionapplic} and in Refs.\
\onlinecite{Wern02b} and \onlinecite{Wern02c}.

Note that the estimated value for Hund's rule coupling\cite{LL00} in
Sr$_2$RuO$_4$ of $J_{\rm H} \approx 0.2 - 0.4$ eV is larger than the
estimate for the spin-orbit coupling\cite{NS00} of $\lambda \approx
0.1$ eV. Consequently the effective model for the magnetic
correlations derived here is applicable even in the presence
corrections from spin-orbit coupling that lifts the degeneracy of the
$d_{zx}$ and $d_{yz}$ orbitals because the larger Hund's rule coupling 
overcompensates the effect.


\section{Inter-plane coupling and band hybridization}\label{sectionhybrid}

The model described in Sec. \ref{sectionzbands} has been based on the
two $d_{zx}$ and $d_{yz}$ bands that are coupled via the on-site
interaction only. In this section we discuss the $d_{xy}$ band as well
as the $d_{zx}$-$d_{yz}$ inter-chain and hybridization terms which are
expected to qualitatively change the low temperature physics of the
bosonized model. Here we discuss the magnitude of the terms and their
expected impact. In Sec.\ \ref{sectionapplic} we then discuss at what
temperature which properties of Sr$_2$RuO$_4$ are determined by a
certain subsystem. 
 
The in-plane resistivity is two to three orders of magnitude smaller
than that along the $c$ axis.\cite{MYH+97} Consistently the dispersion
of the Fermi energy along $c$ is about 1\% of the in plane
dispersion as probed be dHvA measurements.\cite{BJM+00} Band structure
calculations lead to an estimated inter-plane hopping of about 10\% of
the in-plane hopping.\cite{MS97}   

The appropriate inter-plane Hamiltonian with hopping amplitude
$t_\perp$ is 
\begin{equation}\label{Hperp}
H_{\perp}= t_\perp \sum_{\nu,\nu'=x,y}\,\sum_{\bl,\bl',\sigma}
         c^{\dagger}_{{\bl},\nu,\sigma} 
         c^{\phantom{\dagger}}_{{\bl}',\nu,\sigma}\,,
\end{equation}
with only nearest neighbors $\bR_{\bl'}=\bR_\bl+\frac{1}{2}(\pm a,\pm
a, \pm c)^\dagger$. The inter-plane hopping of the $d_{xy}$ band is an 
order of magnitude smaller\cite{BJM+00} as a consequence of the
in-plane geometry of the $d_{xy}$ orbitals\cite{MRS01} and can be 
neglected. Fourier transforming the Fermi operators via
\begin{equation}\label{FTFermi}
c_{{\bl},\nu,\sigma} = \frac{1}{\sqrt{N}}\sum_{\bk} 
 {\rm e}^{i\bk\bR_\bl}
         c_{{\bk},\nu,\sigma} 
\end{equation}
leads to 
\begin{equation}\label{Hperpk}
H_{\perp}= 8 t_\perp \!\sum_{\nu,\sigma,\bk}\!
\cos\frac{a k_x}{2} \cos\frac{a k_y}{2} \cos\frac{c k_z}{2}\,
         c^{\dagger}_{{\bk},\nu,\sigma} 
         c^{\phantom{\dagger}}_{{\bk},\nu,\sigma}\,.
\end{equation}
The in-plane kinetic energies of the  $d_{zx}$ and $d_{yz}$ electrons
are  
\begin{equation}\label{Hinplane}
H_{\nu}=2 t_0 \sum_{\sigma,\bk}
\cos(a k_\nu)\
         c^{\dagger}_{{\bk},\nu,\sigma} 
         c^{\phantom{\dagger}}_{{\bk},\nu,\sigma}\,.
\end{equation}
The total Hamiltonian for the $d_{zx}$ and $d_{yz}$ electrons
$H_x+H_y+H_{\perp}$ can readily be diagonalized and yields the
dispersion for the $\alpha$ and $\beta$ bands as 
\begin{eqnarray}\label{Dispersionab}
E_\bk^{\left({\alpha \atop \beta}\right)} 
&=& E_0 + t_0 \left(\cos a k_x + \cos a k_y\right)
+ 8t_\perp g_{1,\bk} \nonumber\\&&\hspace{12ex}
\pm\frac{1}{2}
\sqrt{t_0^2 g_{0,\bk}^2 + 256 t^2_\perp g^2_{1,\bk}}\ .
\end{eqnarray}
The abbreviations $g_{0,\bk} = \cos(a k_x) - \cos(a k_y)$ and
$g_{1,\bk} = \cos\frac{a k_x}{2} \cos\frac{a k_y}{2} \cos\frac{c
k_z}{2}$ were introduced for lucidity.   

\begin{figure}[bt]
   \epsfxsize=0.48\textwidth
   \epsfclipon
   \centerline{\epsffile{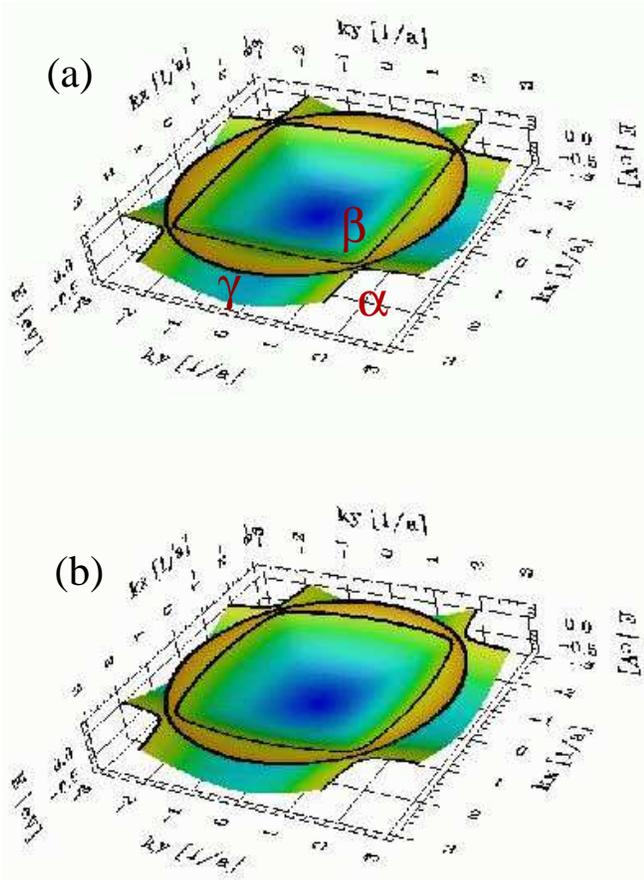}}
   \centerline{\parbox{\textwidth}{\caption{\label{xybands}
   \sl Tight binding model for the three bands that form the Fermi
   surface. (a) Reduced model with $d_{zx}$-$d_{yz}$ hybridization
   trough inter-plane coupling Eq.\ (\protect\ref{Dispersionab}). (b)
   Model including next nearest layer hopping, inter-chain coupling,
   and interaction induce on-site $d_{zx}$-$d_{yz}$ hybridization Eq.\
   (\protect\ref{Dispersiontilde}) for $k_z=0$.}}}      
   \end{figure}

Panel (a) of Fig.\ \ref{xybands} shows the resulting tight binding
bands for $k_z=0$ as a function of $k_x$ and $k_y$. The parameters
are $E_0 = 0.22$ eV, $t_0 = -0.3$ eV, and $t_\perp = -0.02$ eV. The
dispersion of the $d_{xy}$ or $\gamma$ band is given by 
\begin{eqnarray}\label{Dispersiong}
\frac{E_\bk^{(\gamma)}}{\rm eV} 
&=& - 0.39 - 0.54 (\cos ak_x + \cos ak_y )
\nonumber\\&&\hspace{15ex}
- 0.44 \cos ak_x  \cos ak_y\ .
\end{eqnarray}
The term $\sim \cos ak_x  \cos ak_y$ stems from in-plane next nearest
neighbor hopping. The corresponding real space hopping parameters are
$t^{\gamma,\gamma}_{{\bl},{\bl}} = -0.39$ eV, 
$t^{\gamma,\gamma}_{{\bl},{\bl}+\hat{x}} =
t^{\gamma,\gamma}_{{\bl},{\bl}+\hat{y}} = -0.27$ eV, and 
$t^{\gamma,\gamma}_{{\bl},{\bl}+\hat{x}+\hat{y}} = -0.11$ eV. 
The qualitative agreement with ARPES,\cite{DLS+00}
LDA,\cite{MS97} and dHvA\cite{MJD+96,MRS01} results is satisfactory.

In order to obtain a more precise match with the three-dimensional
Fermi surface suggested by dHvA measurements\cite{MJD+96,BJM+00} it is
necessary to extend the dispersion for the $\alpha$ and $\beta$ sheets
in Eq.\ (\ref{Dispersionab}). The non-vanishing coefficient $k_{02}$
in Ref.\ \onlinecite{BJM+00} suggest a term $2t_z \cos ck_z$ 
from next nearest layer hopping. Spin-orbit coupling in the
$d_{zx}$-$d_{yz}$ subsystem leads to a hybridization of the
orbitals.\cite{NS00,EMB02} The fields in the Hamiltonians discussed in
Sec.\ \ref{sectionbose} are describing hybridized $d_{zx}$ and $d_{yz}$
orbitals as a consequence of the on-site interaction.\cite{HCW02} In
the framework of the tight binding model these effects can be modeled
by introducing the on-site hybridization $t_h$. The in-plane dispersion
is also enhanced and can be modeled by extending the diagonal
contributions Eq.\ (\ref{Hinplane}) to include an effective
inter-chain hopping $t_0 \cos ak_\nu \to t_0 \cos ak_\nu + t_i \cos
ak_{\nu'\neq\nu}$. The resulting dispersions are  
\begin{eqnarray}\label{Dispersiontilde}
\tilde{E}_\bk^{\left({\alpha \atop \beta}\right)} 
&=& \tilde{E}_0 + (t_0+t_i) \left(\cos a k_x + \cos a k_y\right) 
+  2t_z \cos ck_z
\nonumber\\&&\hspace{-0ex}
+\ 8t_\perp g_{1,\bk}
\pm\frac{1}{2}
\sqrt{t_0^2 g_{0,\bk}^2 + 256 t^2_\perp g^2_{1,\bk}
 +4 t_h^2}\ .\nonumber\\&&
\end{eqnarray}
An appropriate choice of parameters is $\tilde{E}_0 = -0.29$ eV,
$t_0=0.3$ eV, $t_i=0.03$ eV, $t_z=0.02$ eV, $t_\perp=0.02$ eV, and
$t_h=0.06$ eV. The resulting bands are shown in Fig. \ref{xybands}
panel (b) for $k_z=0$.

The tight binding analysis leads to the following conclusions.

(i) The two- and three-dimensional corrections to the quasi
one-dimensional $d_{zx}$ and $d_{yz}$ bands are of the order of 10\%
or $0.03$ eV leading to the presence of the $\alpha$ and $\beta$
sheets of the Fermi surface. Luttinger liquid behavior should only be
observable at sufficiently high temperatures. For $T > t_{i,z,\perp,h}
\sim 400$ K the out-of-plane transport is incoherent and saturates
while the in-plane resistivity is determined by the crossed sliding
Luttinger liquid with linear temperature dependence. Consequently the
model is consistent with the observed anomalous high temperature
resistivity.\cite{TMNM98} Also consistent is that the quasi-particle
peaks observed in ARPES disappear above $T\sim 160$
K.\cite{WYK+02,VJY+02}

(ii) Since the on-site interactions are an order of magnitude larger
than the tight binding parameters, i.e., $U_{0,1,2}\gg
t_{i,h,\perp,z}$, dominant correlation effects are still determined
or at least influenced by the Hamiltonians (\ref{effH}),
(\ref{Hnullc}), and (\ref{Hintz}) with properly renormalized
parameters. Examples are the magnetic structure factor (Sec.\
\ref{sectionM}), the specific heat (Sec.\ \ref{sectionC}) and the
degenerate superconducting saddle point discussed in Ref.\
\onlinecite{Wern02b}. An account of the temperature dependence of the
corrections due to the hybridization terms $t_{i,h,\perp,z}$ is
given at the beginning of Sec.\ \ref{sectionapplic}.

(iii) As discussed in Ref.\ \onlinecite{Wern02b} the inter-plane
hopping is the important parameter for the mean field superconducting
transition and is estimated to be $t_\perp \sim 20$ meV. The resulting
coefficient $t_\perp^2/\vF\sim 6$ K of the inter-plane pair
hopping term is consistent with the transition temperature of $T_{\rm
c} = 1.5$ K on the mean field level.


\section{Application to S\lowercase{r}$_2$R\lowercase{u}O$_4$}
\label{sectionapplic} 


The model derived in Sec.\ \ref{sectionzbands} readily accounts for
the observed normal phase properties in Sr$_2$RuO$_4$. The most
striking qualitative evidence is the recently discovered scale
invariance of the magnetic structure factor\cite{BSB+02} in agreement
with the implications from the effective one-dimensional model Eq.\
(\ref{effH}). The scale invariance has been observed for
$T=60,110,160$ K while it starts to break down\cite{BSB+02} at $T=10$
K which is in the regime of Fermi liquid behavior.\cite{MYH+97,BJM+00}   

Therefore the hierarchy of the applicability of the model derived in
Sec.\ \ref{sectionzbands} can be summarized as follows. For $T > 400$
K we expect crossed sliding Luttinger liquid
behavior.\cite{MKL01,TMNM98} Curvature corrections to the linearized 
bands of the order of 10\% are conceivable at $T \sim 700$ K. For
$400\ {\rm K} > T > 25\ {\rm K}$ the system gradually crosses over to
the Fermi liquid regime because the various coupling terms discussed
in Sec.\ \ref{sectionhybrid} become relevant at different
temperatures.\cite{VJY+02} The observed scale invariance of the
magnetic excitations suggest that these relevant terms mostly impact
the quasi-two dimensional charge channel given by Eq.\
\ref{Hnullc}. The one- to two-dimensional crossover of the magnetic
subsystem given by Eq.\ \ref{Hnullm} is determined by
$\overline{V}_{\rm m}$. Since the charge and magnetic channels are
coupled via Eq.\ (\ref{Hintz}) the Fermi liquid behavior only is fully
observed in the electronic channel when the quasi one-dimensional 
magnetic fluctuations are frozen out for $T\
\mbox{\raisebox{-3pt}{$\stackrel{\textstyle <}{\sim}$}}\
\overline{V}_{\rm m}$. From the experimentally observed onset of Fermi 
liquid behavior\cite{MYH+97,BJM+00}  we estimate $\overline{V}_{\rm m}
\approx 25$ K. Equivalently, a crossover to non-Fermi liquid behavior
on energy scales $\omega > 2$ meV is expected.

\subsection{Incommensurate magnetism}\label{sectionM}

An important probe for the interaction effects in correlated electron
systems is the magnetic structure factor determined by neutron
scattering. The quasi one-dimensional model derived in Sec.\
\ref{sectiononeeff} accounts for the dominant features of the magnetic
response.
 
\begin{figure}[bt]
   \epsfxsize=0.48\textwidth
   \epsfclipon
   \centerline{\epsffile{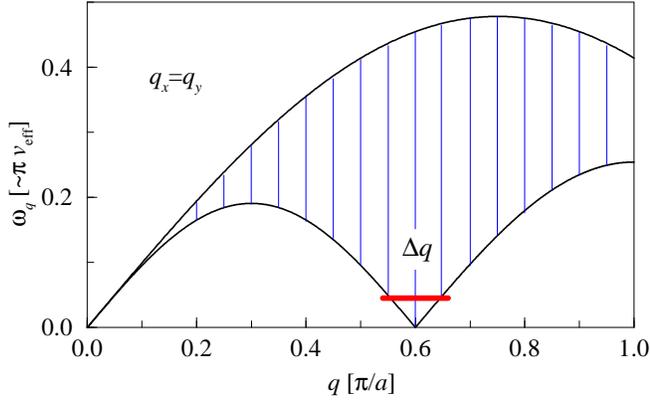}}
   \centerline{\parbox{\textwidth}{\caption{\label{magnspect}
   \sl Sketch of the spectrum of the elementary magnetic excitations
   of a Heisenberg chain in a magnetic field as adapted from Ref.\
   \protect\onlinecite{KM00}. It models the magnetic excitations of
   the Hamiltonian Eq.\ (\protect\ref{effH}) along $q_x=q_y$. The bar
   marks the width of the spectrum $\Delta q$ for a given energy
   transfer.}}}        
   \end{figure}

The bosonization approach correctly describes excitations near the
Fermi surface,\cite{Frad91} i.e., for momentum transfer $q\sim 0$ and 
$q \sim 2k_{\rm F}$. The relevant momentum transfer for
antiferromagnetic magnetic excitations\cite{Schu86} is $q \sim 2k_{\rm
F}$ since the integrated intensity of the structure factor for $q\to
0$ vanishes.\cite{KMB+97} A one-dimensional model analogous to that of
Eq.\ (\ref{Hnullom}) with equivalent bosonized Hamiltonian and
incommensurate back scattering wave vector $q \neq \frac{\pi}{a}$ is
the antiferromagnetic Heisenberg chain in a uniform
field.\cite{FGM+96} We thus expect the effective one-dimensional
Hamiltonian Eq.\ (\ref{Hnullom}) to describe an excitation spectrum as
determined in Ref.\ \onlinecite{KM00} and sketched in Fig.\
\ref{magnspect}. Since Eq.\ (\ref{Hnullom}) is a one-dimensional model
along the diagonal of the basal plane of the unit cell of
Sr$_2$RuO$_4$, since the model is manifestly invariant under the
symmetry transformations $x \leftrightarrow y$ and $y \to -y$, and
since wave vectors are only defined modulus a reciprocal lattice
vector we find gapless magnetic excitations with linear dispersion at
$\bq_{i} = (\pm [\frac {2\pi}{a} - 2 k_{\rm F}], \pm[\frac {2\pi}{a} -
2 k_{\rm F}])^\dagger \approx (\pm 0.6\pi/a,\pm 0.6\pi/a)$ [compare
Fig.\ \ref{xybands} and Refs.\ \onlinecite{MS97,Wern02d}].

The result from conformal field theory for any two point correlation
function\cite{Schu86,GNT98} is valid for sufficiently small
frequencies $\omega$ and momenta $\oq = (q-|\bq_i|)a$ measured with
respect to the back scattering wave vector.
\begin{equation}\label{ChiCFT}
\chi_{\rbx{\rm m}}(\oq,\omega) = \frac{A_x}{T^{2-2x}} 
I_x\left(\frac{\omega-v_{\rm eff}\,\oq}{2\pi T}\right)
I_x\left(\frac{\omega+v_{\rm eff}\,\oq}{2\pi T}\right)\,,
\end{equation}
where
\begin{equation}\label{function}
I_x(k) = \frac{\Gamma(x/2-ik/2)}{\Gamma(1-x/2-ik/2)}\ .
\end{equation}
The value of the scaling dimension $x$ and the excitation velocity
$v_{\rm eff}$ depend on the details of the system. The prefactor $A_x$ 
depends on the scaling dimension. Please refer to Ref.\
\onlinecite{WK01} for details.

The limits of the applicability of the result from conformal field
theory can be understood in the framework of studies of Heisenberg
chains performed in Ref.\ \onlinecite{WK01}. It has been shown that
the effective scaling dimension $x$ is temperature dependent. At or
above temperatures of the order of the excitation velocity, i.e., $T\
\mbox{\raisebox{-3pt}{$\stackrel{\textstyle >} {\sim}$}}\ v_{\rm eff}$
we expect the scaling dimension to attain the non-interacting limit,
i.e., $x\to 1$. At energy transfers of the order of and above the
excitation velocity lattice corrections become relevant. Similar
arguments hold for the momentum transfer. At finite temperatures and
finite energy transfer the effects combine and the 
range of validity of Eq.\ \ref{ChiCFT} can roughly be estimated as
$\sqrt{T^2 + \omega^2 +(v_{\rm eff}\,\oq)^2} \le v_{\rm eff}/2$.

The following experimental observations can be understood within the
framework of the outlined analogies.

(i) The imaginary part of the magnetic correlation function at small
energy transfer is strongly peaked at $\bq_{i}$ and flat
elsewhere.\cite{Schu86,KMB+97} This is in perfect qualitative
agreement with the magnetic structure factor $S(\bq,\omega) \sim {\rm
  Im}\,\chi_{\protect\rbx{\rm m}}(\bq,\omega)$ determined via
inelastic neutron scattering.\cite{SBB+99,BSB+02}  

(ii) The magnetic correlation function Eq.\ (\ref{ChiCFT}) is scale
invariant. The scale invariance has been observed experimentally
outside the Fermi liquid regime\cite{BSB+02} and suggests values of
$1/2 \le x \le 5/8$ for Sr$_2$RuO$_4$. Note that these values of 
$x$ describe a $XXZ$ Heisenberg chain near the isotropic
point,\cite{FLS97} i.e., $J\
\mbox{\raisebox{-3pt}{$\stackrel{\textstyle >} {\sim}$}}\  J_{z}$ for
in-plane ($J$) and out-of-plane ($J_{z}$) magnetic
coupling,\cite{WK01} and are thus in quantitative agreement with the
intermediate coupling regime assumed for Sr$_2$RuO$_4$ within this
approach.

\begin{figure}[bt]
   \epsfxsize=0.48\textwidth
   \epsfclipon
   \centerline{\epsffile{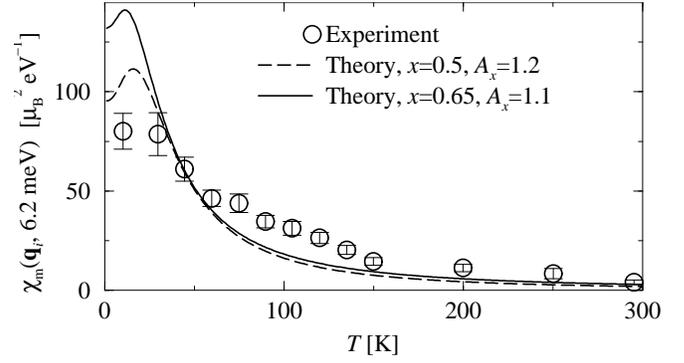}}
   \centerline{\parbox{\textwidth}{\caption{\label{ChiofT}
   \sl Plot of $\chi_{\protect\rbx{\rm m}}(0,6.2\ {\rm meV})$ from
   Eq.\ (\protect\ref{ChiCFT}) for $x=0.5$ (broken line) and $x=0.65$
   (full line) in comparison with experimental\protect\cite{SBB+99} 
   results. Good agreement of the fits for fixed $x$ can only be
   expected in the temperature interval $2\overline{V}_{\rm m} \le T
   \ll v_{\rm eff}$.}}}         
   \end{figure}

(iii) Figure \ref{ChiofT} shows the temperature dependence of ${\rm Im}\, 
\chi_{\rbx{\rm m}}(0,6.2\ {\rm meV})$ from Eq.\ (\ref{ChiCFT}) for  
$x=0.5$ (broken line) and $x=0.65$ (full line) in comparison with
experimental\cite{SBB+99} results. The fits are only valid outside the
Fermi liquid regime, i.e., for $T > 25$ K. Together with the
limitations of the applicability of Eq.\ (\ref{ChiCFT}) discussed
above the agreement of the fits for fixed $x$ with the experimental
data can only be expected in a small temperature interval of
$2\overline{V}_{\rm m} \le T \ll v_{\rm eff}$. Note the relatively
large energy transfer of $\omega = 6.2\ \mbox{\rm meV} = 73$
K. Consequently a discrimination between $x=0.5$ and $x=0.65$ is not
conclusive. The prefactors $A_{0.5} = 1.2$ and $A_{0.65} = 1.1$ are of
the correct order of magnitude. Neutron scattering results at lower
energy transfer are desirable.

\begin{figure}[bt]
   \epsfxsize=0.48\textwidth
   \epsfclipon
   \centerline{\epsffile{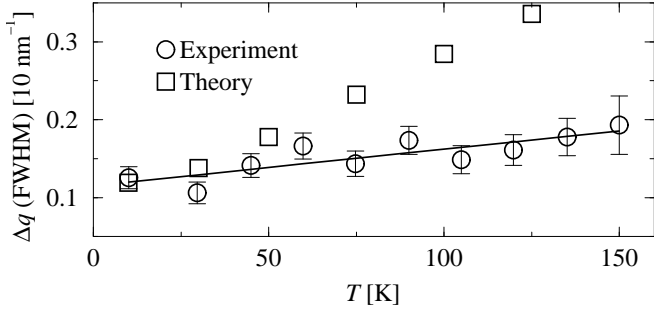}}
   \centerline{\parbox{\textwidth}{\caption{\label{DqofT}
   \sl Full width at half maximum as determined by neutron
   scattering\protect\cite{SBB+99} (circles) and as obtained from Eq.\
   (\protect\ref{ChiCFT}) with $\omega=6.2$ meV, $x=0.5$, and $v_{\rm
   eff} = 350\ {\rm K}$ (squares). The theoretical values for $T\ge
   75$ K are less reliable since $T$ and $\oq$ become too large. The
   line is a guide to the eye.}}}        
   \end{figure}

(iv) The experimental results show a width of the magnetic peaks which
is only weakly temperature dependent\cite{SBB+99} as shown by the
circles in Fig.\ \ref{DqofT}. In the model presented here the finite
width of the dynamic magnetic correlations follows out of the
dispersion of the lower bounds of the excitation continuum as
indicated by the bar near $q \sim \frac{2\pi}{a} - 2k_{\rm F}$ in
Fig.\ \ref{magnspect} with $\Delta q \sim \frac{2\omega}{v_{\rm eff}
  a}$. 

The temperature dependence of the excitation velocity can be estimated
by considering the magnetic correlation function ${\rm Im}\,
\chi_{\rbx{\rm m}}(\oq,6.2\ {\rm meV})$ from Eq.\ (\ref{ChiCFT}) as a 
function of $\oq$ and determining the full width at half maximum
(FWHM). The results are shown for $x=0.5$ and $v_{\rm eff} = 350\ {\rm 
  K}$ as the squares in  Fig.\ \ref{DqofT}. Consistent with the
expected limits of validity of Eq.\ (\ref{ChiCFT}) and the temperature
dependence of the peak intensity discussed in paragraph (iii) the
values of $\Delta q$ are in good agreement with the experimental data
only for $T < 75$ K. Note that for $T > 75$ K also the relevant
values of $\oq$ become large. For $x=0.65$ the results only differ of
the order of 10\% and do not allow for any quantitative
discrimination. We conclude that the magnetic energy scale is given by
$v_{\rm eff} \sim 10^2\ {\rm K}$.\cite{Mrenormquote}

\begin{figure}[bt]
   \epsfxsize=0.48\textwidth
   \epsfclipon
   \centerline{\epsffile{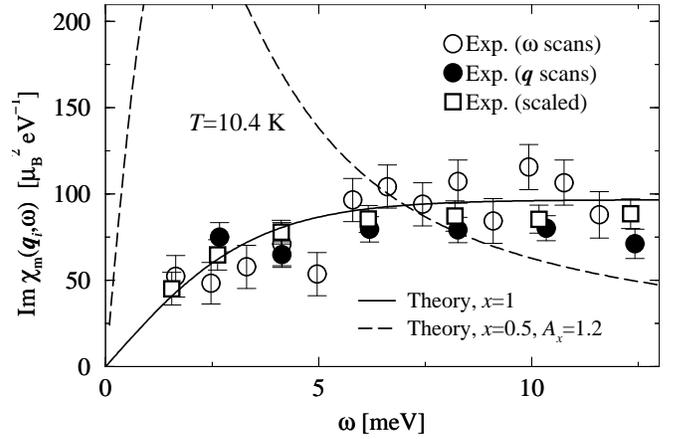}}
   \centerline{\parbox{\textwidth}{\caption{\label{Chiofw}
   \sl Plot of ${\rm Im}\,\chi_{\protect\rbx{\rm m}}(0,\omega)|_{T =
   10.4\ {\rm K}}$ from
   Eq.\ (\protect\ref{ChiCFT}) for $x=1$ (full line) and
   $x=0.65$ (broken line) in comparison with
   experimental results for frequency scans (open
   circles\protect\cite{SBB+99} and open
   squares\protect\cite{BSB+02}) and momentum scans (full
   circles\protect\cite{SBB+99}). Since at $T \sim 10$ K the system is
   in the Fermi liquid regime and for $\omega > 2$ meV only gradually
   crosses over to the conformally invariant regime the applicability
   of Eq.\ (\protect\ref{ChiCFT}) is not obvious (see text).}}}          
   \end{figure}

(v) The symbols in Fig.\ \ref{Chiofw} show the energy dependence of 
${\rm Im}\, \chi_{\rbx{\rm m}}(0,\omega)$ at $T=10.4$ K from neutron
scattering measurements. Open and full circles are from Ref.\
\onlinecite{SBB+99} for energy and $\bq$ scans, respectively. Squares
from Ref.\ \onlinecite{BSB+02} are scaled since no absolute scale is
given. The data are in qualitative agreement with the presence of an
excitation continuum.

Since for $T=10.4$ K the system is in the Fermi liquid regime
the applicability of Eq.\ (\ref{ChiCFT}) is not obvious. The
experimental data can be fitted with the (renormalized)
non-interacting case, where\cite{FLS97} $x=1$ and ${\rm Im}\,
\chi_{\rbx{\rm m}}(0,\omega) \sim v_{\rm eff}^{-1} \tanh[\omega/(4T)]$
as shown by the full line in Fig.\ \ref{Chiofw}. The amplitude of the
fit has been chosen as $v_{\rm eff} = 120$ K consistent with $v_{\rm
  eff}\sim 10^2$ K. The good agreement of the fit is likely to be
accidental since for larger frequencies $\omega > 2$ meV the system
should gradually cross over from Fermi liquid to conformal
behavior. For $\omega \sim v_{\rm eff}$ effects from the upper
continuum limit become relevant\cite{WK01} voiding the direct
applicability of Eq.\ (\ref{ChiCFT}) for $\omega\
\mbox{\raisebox{-3pt}{$\stackrel{\textstyle >}{\sim}$}}\ 10$ meV. The
broken line in Fig.\ \ref{Chiofw} shows the result for the interacting 
case with $x=0.5$ and $A_{0.5}=1.2$ as determined under (iii) for
comparison.

Measurements of the low energy dynamical structure factor outside the
Fermi liquid regime at $30\ {\rm K} \le T \le 50 $ K are desirable to
test the theory presented here in its range of applicability and allow
conclusive comparison with perturbative\cite{EMJB02} results.

(vi) The presence of quasi one-dimensional correlations along the
system diagonals finds further experimental support in the
non-analytic angular dependence of the in-plane upper critical
fields.\cite{Wern02c}

In conclusion the functional dependence of the dominant magnetic
correlations anticipated from conformal field theory as given by Eq.\ 
(\ref{ChiCFT}) describes the experimental data satisfactorily within
the framework of the expected applicability of the theory. The
consistency of the results suggests that we were able to extract a
reliable energy scale for the effective magnetic correlations.

\subsection{Comparison with RPA}\label{sectionRPA}

The magnetic correlations in Sr$_2$RuO$_4$ have been widely studied
theoretically\cite{MS99,BSB+02,LL00,EMJB02,EMB02,MTG01,BMJ+02,OO02a}
using perturbative approaches such as the random phase approximation
(RPA). The perturbative approaches cannot account for the low
dimensional quantum fluctuations and it is instructive to discuss the
resulting limits of their applicability. To this end we have performed
a RPA analysis of the magnetic structure factor. The interaction is
included in the dynamical correlation functions via\cite{Maha90} 
\begin{equation}\label{chiRPA}
\chi_{\rbx{\rm RPA}}^{\nu,\nu'}(\bq,\omega) = 
\chi_{\rbx{0}}^{\nu,\nu'}(\bq,\omega) \left[
            1 - U^{\nu,\nu'}_{\sigma,\sigma'}\ 
       \chi_{\rbx{0}}^{\nu,\nu'}(\bq,\omega) \right]^{-1}\,.
\end{equation}
The bare susceptibilities $\chi_{\rbx{0}}^{\nu,\nu'}(\bq,\omega)$ are
the Fourier transforms of the real time spin-spin correlation function
$-i \theta \langle \bS^{\nu}_{\bq}(t) \bS^{\nu'}_{-\bq}(0)
\rangle$ and are determined with respect to the tight binding model
discussed in Sec.\ \ref{sectionhybrid}. The spin operators
$\bS^{\nu}_{\bq}(t)$ act on electrons with orbital index
$\nu,\nu'\in\{x,y,\gamma\}$. In the absence of Hund's rule coupling
the interactions are $U^{\nu=\nu'}_{\sigma\neq\sigma'} = U_0$ and
$U^{\nu\neq\nu'}_{\sigma,\sigma'} = U_1 = U_2$. For $U_1\neq U_2$ the
correlation function contributions to
$\chi_{\rbx{\sigma,\sigma'}}^{\nu,\nu'}$ are also anisotropic in the
spin Hilbert space.\cite{NS00}  A very recent
approach\cite{OO02a} including Hund's rule coupling suggests the
stabilization of a chiral magnetic state. The thermodynamic
expectation values are determined via the interaction free Hamiltonian
with tight binding bands as shown in Fig.\ \ref{xybands} and given in
Eqs.\ (\ref{Dispersiong}) and (\ref{Dispersiontilde}).

\begin{figure}[bt]
   \epsfxsize=0.5\textwidth
   \epsfclipon
   \centerline{\epsffile{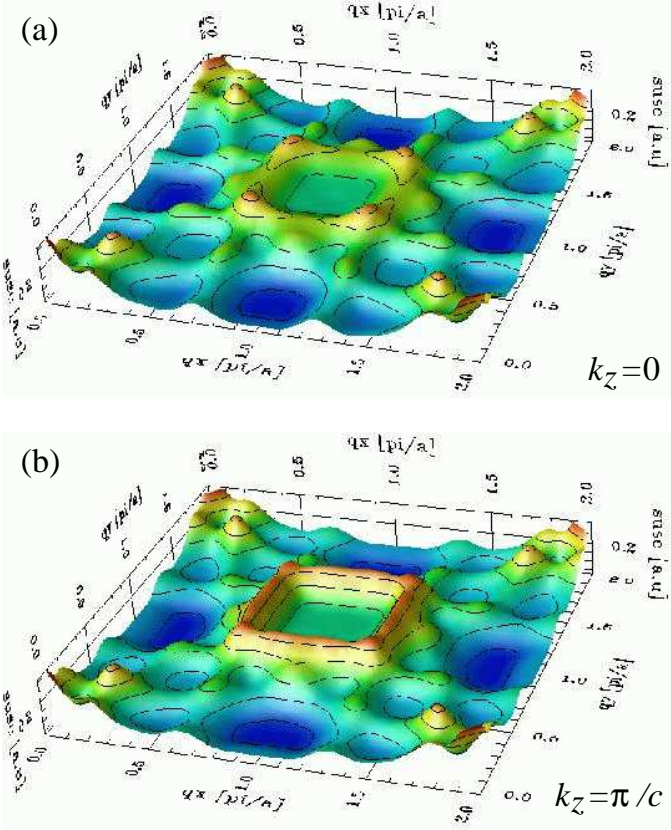}}
   \centerline{\parbox{\textwidth}{\caption{\label{chiRPAps}
   \sl Tr ${\rm Im}\, \chi_{\protect\rbx{\rm
   RPA}}^{\nu,\nu'}(\bq,\omega)$ convoluted with the experimental $q$
   resolution. Parameters are $\omega=6.2$ meV, $U_0 = 0.2$ eV, $U_1 =
   0.1$ eV, $U^{x,\gamma}_{\sigma,\sigma'} =
   U^{y,\gamma}_{\sigma,\sigma'} = 0$ and dispersions Eqs.\ 
   (\protect\ref{Dispersiong}) and
   (\protect\ref{Dispersiontilde}). Panel (a) $k_z=0$ and (b) 
   $k_z=\pi/c$ show the sensitivity of the RPA approach to small
   changes in the parameters.}}}       
   \end{figure}

Figure \ref{chiRPAps} shows ${\rm Im}\, \chi_{\rbx{\rm RPA}}^{\rm
tot}(\bq,\omega) = {\rm Tr}\, {\rm Im}\, \chi_{\rbx{\rm
RPA}}^{\nu,\nu'}(\bq,\omega)$ for $\omega=6.2$ meV convoluted with the
resolution $\delta q\approx 0.1\pi/a$ from neutron scattering
experiments\cite{SBB+99} (a) for $k_z=0$ and (b) for $k_z=\pi/c$. The
interaction parameters here are $U_0 = 0.2$ eV, $U_1 = 0.1$ eV, and
$U^{x,\gamma}_{\sigma,\sigma'} = U^{y,\gamma}_{\sigma,\sigma'} =
0$. This neglects the hybridization of the $d_{xy}$ band with the
$d_{zx}$ and $d_{yz}$ bands which has a quantitative effect on
$\chi_{\rbx{\rm RPA}}^{\gamma,\gamma}(\bq,\omega)$,\cite{EMJB02} but
not so much on the total correlation function. The features discussed
in Ref.\ \onlinecite{EMJB02} are reproduced albeit with different
weight.  

Panel (a) of Fig.\ \ref{chiRPAps} shows the correlations in the plane
through the $\Gamma$ point of the Brillouin zone ($k_z=0$) while the
correlations in (b) lie in the plane through the mid point of the line
$\overline{\Gamma Z}$ ($k_z=\pi/c$).\cite{DLS+00} The difference of
the two shows the sensitivity of the RPA approach to small changes in
the parameters.    

The total correlation function in RPA in Fig.\ \ref{chiRPAps} clearly
shows the structures of the bands dispersing in the plane. Many of
these structures have not been observed
experimentally.\cite{BSB+02} Moreover, the parameters have to be
fine-tuned in the RPA approach close to a phase
transition.\cite{SBB+99} Both effects can be understood as
consequences of the underestimation of quantum fluctuations.

Quantum fluctuations in low dimensions suppress long range order even
if the value of the interaction is large enough to give a finite
temperature phase transition in RPA.\cite{MW66} More specifically, for
the one-dimensional case---which is discussed closely in the review by 
S{\'o}lyom\cite{Soly79}---the relevance of back and umklapp scattering
terms is overestimated by perturbative approaches such as
RPA. Instead, their relevance has to be determined non-perturbatively,
namely via renormalization group (RG) studies.

Consequently the two-dimensional RPA approach tends to underestimate
the one-dimensional correlations since the renormalization of the
excitation velocities can only be modeled indirectly by the
interaction strength while the relative size of the two-dimensional
features tends to be overestimated. Because of the large parameter space
of the RPA approach and the sensitivity to details in the
band structure it is still possible to model the low temperature
magnetic structure factor in Sr$_2$RuO$_4$ rather
accurately.\cite{BSB+02} The weakness of the RPA approach becomes
apparent through the fact that the description of different properties 
of the material requires different choices of parameter
sets.\cite{EMJB02,MTG01} 

The present approach allows to include the RG results from the
literature\cite{SCP98} since the back and umklapp scattering terms are
given explicitly by Eq.\ (\ref{Hintz}) as discussed in Sec.\
\ref{sectionbose}. The quantum fluctuations and interaction effects
are included by the renormalization of parameters such as $v_{\rm
  eff}$ and their impact becomes apparent through the small scaling
dimension $x$ discussed in Sec.\ \ref{sectionM}.  

In conclusion the incommensurate magnetic fluctuations in
Sr$_2$RuO$_4$ are best described via the quasi one-dimensional
correlations from Eq.\ (\ref{effH}) with a spectrum as sketched in
Fig.\ \ref{magnspect}.\cite{Wern02d} The two-dimensional correlations 
yield an enhanced but relatively homogeneous background with only
small additional structures.\cite{BSB+02} Note that the
two-dimensional magnetic correlations beyond the diagonals are given
within the present approach by Eq.\ \ref{Hnullm}.


\subsection{Effective electronic masses}\label{sectionmasses}

In the following we will discuss effective electronic masses observed
in the Fermi liquid regime of Sr$_2$RuO$_4$. Clearly, a system that
shows scale invariant correlations down to 25 K must show a different
renormalization of the Fermi liquid parameters than a system that is a
Fermi liquid at all temperatures. Consequently it is quite natural to
estimate that difference phenomenologically by considering that there
is a very strong reduction of the effective excitation velocity of
$v_{\rm F}/v_{\rm eff} \sim 20 - 60$ in the magnetic sector along the
diagonals of the basal plane due to interaction effects as discussed
in Sec.\ \ref{sectionM}. In the Fermi liquid regime this reduction
cannot be calculated directly via the results from conformal field
theory (Sec.\ \ref{sectionM}) because of the two-dimensional coupling
$\overline{V}_{\rm m}$ (Sec.\ \ref{sectionbose}). It is yet reasonable
to assume that the reduction of the excitation velocity in the
magnetic channel along the diagonals remains much larger than the mere
factor of 2 (see below) obtained in perturbative\cite{LL00}
approaches. With this assumption the different electronic masses can
be modeled.

\subsubsection{Cyclotron mass}\label{sectionCyclo}

The cyclotron mass $m_{\rm c}$ determines the cyclotron frequency and
has been measured in dHvA experiments\cite{BJM+00,BMJ+02} to be
enhanced with respect to the bare (non-interacting) band mass $m_{\rm
  b}$ on all three Fermi surfaces by the same amount, namely $m_{\rm 
  c}^{(\alpha,\beta,\gamma)} / m_{\rm b}^{(\alpha,\beta,\gamma)}
\approx 2$. Since the cyclotron motion does not involve magnetic
excitations the cyclotron mass enhancement due to interactions is
determined in the present approach by the quasi two-dimensional
Hamiltonian of the charge channel Eq.\ (\ref{Hnullc}) in the subsystem
of the $d_{zx}$ and $d_{yz}$ electrons. Perturbative approaches with
an appropriate choice of parameters\cite{LL00} also yield an
electronic mass enhancement factor of 2 for all
bands. ARPES measurements are consistent with this
result.\cite{PSKT98,DLS+00}

\subsubsection{Specific heat}\label{sectionC}

The specific heat of an interacting system can be determined from the
specific heat of the non interacting system via the renormalization of
the thermodynamic mass\cite{BMJ+02} $m^*$ with respect to the bare band
mass $m_{\rm b}$ or, equivalently, by the renormalized excitation
velocities\cite{SCP98} $v_{\rm F}^*$ or $v_{\rm eff}$ with respect to
the bare band Fermi velocity $v_{\rm F}$.

The specific heat of the $d_{zx}$-$d_{yz}$ subsystem $C_{\rm z}$
consists of the two-dimensional contributions from the spin and charge
channels [Eqs.\ (\ref{Hnullc}) and (\ref{Hnullm})] and the
``one-dimensional'' magnetic part along the diagonals [Sec.\
\ref{sectionRPA}]. The renormalized velocity of the two-dimensional
contributions is given through the cyclotron mass enhancement as
$v_{\rm F} / v_{\rm F,2D}^* \approx 2$. The quasi one-dimensional
magnetic correlations with excitation velocity $v_{\rm eff}$ are only
present along the diagonals of the Brillouin zone. Consequently their
contribution to the total specific heat must be weighed with respect
to a system with one-dimensional magnetic correlations throughout the
entire Brillouin zone. The normalized width of the magnetic peaks
$\frac{a}{2\pi} \Delta q\sim 0.07$ discussed in Sec.\ \ref{sectionM}
allow for an estimate. Depending on whether one assumes quasi
one-dimensional correlations in the vicinity of the positions of the
incommensurate fluctuations $\bq_i$ only or along the whole diagonals
weighing factors of $0.02 \le w \le 0.2$ are reasonable. The weight of
the two-dimensional magnetic contribution then is $1-w$.

Together the contributions yield the specific heat of the
$d_{zx}$-$d_{yz}$ subsystem as\cite{SCP98}  
\begin{equation}\label{Cz}
\frac{C_{\rm z}}{T} = \gamma_{\rm z}\, \gamma_0 \approx 
\frac{\vF}{2}\left(\frac{1}{v_{\rm F,2D}^*} +
     \frac{1-w}{v_{\rm F,2D}^*} + 
     \frac{w}{v_{\rm eff}}
 \right) \ \gamma_0\ .
\end{equation}
The value of $\gamma_{\rm z} \approx (m^*_{\alpha} / m_{\rm
b}^{(\alpha)} + 2 m^*_{\beta} / m_{\rm b}^{(\beta)})/3 \approx 3.4$
can be estimated from a weighed average of the thermodynamic masses
$m^*_{\alpha} / m_{\rm b}^{(\alpha)} \approx 3.1$ and $m^*_{\beta} /
m_{\rm b}^{(\beta)} \approx 3.5$ measured via the dHvA
effect.\cite{BMJ+02} Then the ``one-dimensional'' magnetic
contribution is determined as $\frac{w\,\vF}{2\,v_{\rm eff}} \approx
1.4 + w$. Using the value of $v_{\rm eff} = 350$ K considered in Sec.\
\ref{sectionM} (iv) and the bare Fermi velocity\cite{MPS00}
$\vF\approx 0.7$ eV then gives $w = 0.13$. 

It must be pointed out that the two-dimensional coupling $\sim
\overline{V}_{\rm m} \ll V_{\rm m}$ is likely to increase the
effective velocity $v_{\rm eff}$ in the Fermi liquid regime through
the reduction of the correlation effects. This increase is compensated
by an increase of the relevant phase space in the Brillouin zone
determined by $w$. The phenomenological result that the strong
low-dimensional magnetic correlations with $v_{\rm eff} \ll v_{\rm
  F,2D}^*$ enhance the specific heat beyond the value obtained via
perturbative approaches\cite{LL00} remains unaltered.  

For non-interacting electrons in the two quasi one-dimensional bands
under consideration the coefficient $\gamma_0\approx 4.3\ \frac{\rm
  mJ}{\rm K^2 mol}$ results in a specific heat contribution of
$\frac{C_{\rm z}}{T} \approx 15\ \frac{\rm mJ}{\rm K^2 mol}$ that
accounts for about 40\% of the experimentally observed value of
$\gamma_{\rm tot} = 40 \pm 2\ \frac{\rm mJ}{\rm K^2
  mol}$.\cite{NMF+98,NMM99,NMM00,BMJ+02}

Hydrostatic pressure increases the in-plane single particle hopping
which decreases the relative interaction strength and renders the
$d_{zx}$-$d_{yz}$ subsystem more two-dimensional. Consequently the
``one-dimensional'' magnetic contribution in Eq.\ \ref{Cz} is
decreased yielding a natural explication for the observed reduction of
the thermodynamic masses upon application of hydrostatic
pressure\cite{BMJ+02} within the present model.

The thermodynamic mass of the $d_{xy}$ electrons is enhanced by a
factor of $m^*_{\gamma} / m_{\rm b}^{(\gamma)} \sim 5.5$ with respect 
to the bare band mass.\cite{BMJ+02} The renormalized Fermi velocity
from ARPES\cite{PSKT98,DLS+00} or perturbative approaches\cite{LL00}
only accounts for a factor $\sim 2$. The interaction between the
$d_{xy}$ band and the $d_{zx}$-$d_{yz}$ system accounts for a part of
the missing enhancement through coupling to the one-dimensional
correlations in the magnetic channel. Another possible contribution
comes from the proximity of the $d_{xy}$ band to the van Hove
singularity at the $M$ point of the Brillouin zone.\cite{LL00} Nesting
effects\cite{EMJB02,EMB02} yield an additional enhanced magnetic
contribution to the specific heat.

\subsubsection{Static susceptibility and Wilson ratio}\label{sectionW}

Following the argumentation of the specific heat the contributions to
the uniform static magnetic susceptibility of the $d_{zx}$-$d_{yz}$
subsystem also consist of a two- and one-dimensional part. They are
given with respect to the static magnetic susceptibility of the
non-interacting system $\chi_{\rbx{0}}$ as\cite{SCP98} 
\begin{equation}\label{chistat}
\chi_{\rbx{\rm z}} \approx \chi_{\rbx{0}}
   \left(\frac{v_{\rm F} (1-w)}{v_{\rm F,2D}^*} +
         \frac{w\,\vF}{v_{\rm eff}}\right)\approx 4.3\
    \chi_{\rbx{0}}\,.
\end{equation}
All parameters have been fixed previously. Note that here the relative
contribution of the ``one-dimensional'' subsystem is roughly twice
that of the specific heat since $\gamma_0$ includes both magnetic and
charge degrees of freedom while $\chi_{\rbx{0}}$ only accounts for the 
magnetic correlations.

The enhancement of the susceptibility is in reasonable agreement with
the relative spin-mass enhancement of the $\alpha$ and $\beta$ sheets
measured via the dHvA effect\cite{BJM+00,BMJ+02} as $m^*_{\alpha,\rm susc} /
m_{\rm b}^{(\alpha)} \approx 3.7$ and $m^*_{\beta,\rm susc} / m_{\rm
  b}^{(\beta)} \approx 4.3$. Moreover, since the model derived in
Sec. \ref{sectionbose} predicts the quasi one-dimensional magnetic
correlations only in the basal plane of the tetragonal lattice we
expect a magnetic mass enhancement that depends on the position $k_z$
on the Fermi surface with maxima at $k_{z,{\rm max}}(n) = 2\pi n/c$ as
observed\cite{BMJ+02} experimentally.

The ratio of the  uniform static magnetic susceptibility and the
specific heat coefficient has been determined experimentally as
$R_{\rm W} = (\pi^2\chi)/(3\gamma_{\rm tot}) \approx
1.4$.\cite{IKA+97,MYH+97} The value of $R_{\rm W,z} \approx 1.3$
obtained here for the $d_{zx}$-$d_{yz}$ subsystem is in good
agreement. The enhancement of the specific heat has indeed the same
origin as the enhancement of the susceptibility as already concluded
in Ref.\ \onlinecite{MHY+94} which can be identified in the
$d_{zx}$-$d_{yz}$ subsystem as the quasi one-dimensional magnetic
correlations.

An interesting experimental question that arises out of the discussion
above is whether the photo electrons carry signatures of the enhanced
magnetic correlations along the diagonals. To obtain an answer ARPES
data for the $\alpha$ and $\beta$ sheets need to be analyzed in detail
along $\overline{\Gamma X}$ in comparison with $\overline{\Gamma M}$
and $\overline{M X}$. Using the value extracted in Sec.\
\ref{sectionM} (iv) the expected energy scale is $v_{\rm eff} \sim 30$
meV.  

\section{Conclusions}

The dominantly one-dimensional kinetic energy of the electrons in the
$d_{zx}$ and $d_{yz}$ orbitals allows to bosonize their
Hamiltonian. In the presence of interaction this leads to an effective
two-dimensional model. The degrees of freedom can be parameterized in
terms of the four spin, charge, flavor, and spin-flavor fields. 

The presence of hybridization and corrections to the one-dimensional
kinetic energy make the observation of properties of sliding Luttinger
liquids likely only at or above room temperature. This is consistent
with the observed linear temperature dependence of the resistivity at
high temperatures.

In the magnetic sector described by the spin and spin-flavor fields
the interaction at intermediate coupling leads to a quasi
one-dimensional model along the diagonals of the basal plane of the
Brillouin zone. The resulting spectrum of elementary excitations 
accounts for the enhanced dynamical magnetic susceptibility at
$\bq_i=(\pm 2 k_{\rm F},\pm 2 k_{\rm F})$ and the weak temperature
dependence of the $q$ width. The one-dimensional spectrum leads to a
conformally invariant formulation of the magnetic structure factor
consistent with the observed scale invariance. The scaling dimension
is consistent with the intermediate coupling regime. The observed
excitation continuum and temperature dependence of the peak width are
consistent with a magnetic energy scale of $v_{\rm eff} \sim 10^2$
K. The additional two-dimensional correlations are more homogeneous
than  predicted by RPA because of quantum fluctuations. 

The effective thermodynamic mass enhancement together with the value
of the specific heat coefficient is a superposition of two-dimensional
effects as observed in ARPES and the ``quasi one-dimensional''
correlations of the magnetic channel of the $d_{zx}$-$d_{yz}$
subsystem. The enhancement of the static susceptibility has the same
origin. The contribution of the quasi one-dimensional magnetic
subsystem to the specific heat of the $d_{zx}$-$d_{yz}$ subsystem is
about 44\% and about 70\% to the static magnetic susceptibility of the
$d_{zx}$-$d_{yz}$ subsystem.


\section*{Acknowledgments}

We are thankful to C.\ Bergemann, M.\ Braden, S.\ T.\ Carr, N.\ Dietz,
K.\ Kikoin, M.\ S.\ Laad, S.\ Sachdev, G.\ Schneider, A.\ M.\ Tsvelik,
T.\ Valla, W.\ Weber, and M.\ Weinert for instructive and stimulating
discussions. The work was supported by DOE contract number
DE-AC02-98CH10886.


\begin{thebibliography}{10}
\bibitem{Stablequote}
At this point it should be stressed that the tetragonal structure of
  Sr$_2$RuO$_4$ has been found to be very stable (c.f.\ Ref.\
  \onlinecite{BRN+98}). It is thus admissible to employ the symmetry relations
  that follow from the tetragonal point group at all temperatures.
\bibitem{Mrenormquote}
The size of the renormalization of $v_{\rm eff} \sim 10^2$ K with respect to
  $v_{\rm F} \sim 10^4$ K is reasonable. The magnetic energy scales in
  correlated electron systems range from 10 K to 10$^3$ K while the electronic
  energy scales are of the order of 1 eV yielding energy ratios of 10 to
  10$^3$.
\end{thebibliography}
\end{document}